\documentclass{jpsj3}
\usepackage{txfonts}
\usepackage{color}

\title{Photoinduced High-Frequency Charge Oscillations in Dimerized Systems}

\author{Kenji Yonemitsu\thanks{E-mail: kxy@phys.chuo-u.ac.jp}}
\inst{Department of Physics, Chuo University, Bunkyo, Tokyo 112-8551, Japan} %\\

\abst{ Photoinduced charge dynamics in dimerized systems is studied on the basis of the exact diagonalization method and the time-dependent Schr\"odinger equation for a one-dimensional spinless-fermion model at half filling and a two-dimensional model for $\kappa$-(bis[ethylenedithio]tetrathiafulvalene)$_2$X [$\kappa$-(BEDT-TTF)$_2$X] at three-quarter filling. After the application of a one-cycle pulse of a specifically polarized electric field, the charge densities at half of the sites of the system oscillate in the same phase and those at the other half oscillate in the opposite phase. For weak fields, the Fourier transform of the time profile of the charge density at any site after photoexcitation has peaks for finite-sized systems that correspond to those of the steady-state optical conductivity spectrum. For strong fields, these peaks are suppressed and a new peak appears on the high-energy side, that is, the charge densities mainly oscillate with a single frequency, although the oscillation is eventually damped. In the two-dimensional case without intersite repulsion and in the one-dimensional case, this frequency corresponds to charge-transfer processes by which all the bonds connecting the two classes of sites are exploited. Thus, this oscillation behaves as an electronic breathing mode. The relevance of the new peak to a recently found reflectivity peak in $\kappa$-(BEDT-TTF)$_2$X after photoexcitation is discussed. }

\begin{document}
\maketitle

\section{Introduction}
Photoinduced dynamics and phase transitions in itinerant electron systems \cite{koshigono_jpsj06,yonemitsu_pr08,nicoletti_aop16,giannetti_aip16} have received renewed interest with the observation of strengthened or weakened orders and even transient suppression of charge motion,\cite{ishikawa_ncomms14,fukaya_ncomms15,kawakami_prb17} enhancing the prospect for controlling the electronic phase.\cite{mor_prl17,tanaka_prb18} Here, pictures of photoinduced states are not conventional ones obtained after the absorption of photons but rather electrons directly and coherently driven by an oscillating electric field.\cite{kawakami_prl10,matsubara_prb14} In this context, the concept of dynamical localization may play an important role, although it is basically applicable to systems that are driven by continuous waves.\cite{dunlap_prb86,grossmann_prl91,kayanuma_pra94,ono_prb17} Even after a pulse excitation, states induced by a strong electric field have been shown to be similar to states expected for dynamical localization.\cite{ishikawa_ncomms14,kawakami_prb17,naitoh_prb16,yonemitsu_jpsj17a} 

Dynamical localization describes the long-time behavior obtained by time-averaging. Negative-temperature states and inverted interactions have also been discussed by time-averaging after photoexcitation.\cite{tsuji_prl11,tsuji_prb12,yonemitsu_jpsj15,yanagiya_jpsj15} For the long-time behavior, continuous-wave- and pulse-induced phenomena have been compared in a quantitative manner from a broad perspective\cite{nishioka_jpsj14,ono_prb16,yonemitsu_jpsj17b}. However, the picture for short-time behavior has not been discussed in a systematic manner. Thus, it is desirable to present a concrete example. 

Quite recently, a new reflectivity peak has been discovered in photoexcited $\kappa$-(bis[ethylenedithio]tetrathiafulvalene)$_2$Cu[N(CN)$_2$]Br [$\kappa$-(BEDT-TTF)$_2$Cu[N(CN)$_2$]Br] on the high-energy side of the main reflectivity spectrum.\cite{iwai_unpub} This unprecedented peak is narrow. Its energy is independent of the excitation strength and it survives for a while after photoexcitation; thus, it is not due to the optical Stark effect. The associated charge oscillation has been shown to be enhanced near criticality in the ``pressure''-temperature phase diagram.\cite{kagawa_n05} A mechanism of the emergence of such a high-energy peak is theoretically studied using the exact diagonalization method for small clusters in this paper. Thus, the influence of criticality is beyond the scope of this paper. 

The object material is one of the $\kappa$-(BEDT-TTF)$_2$X, which are quasi-two-dimensional three-quarter-filled dimerized organic conductors. Photoinduced insulator-metal transitions are known to take place in these materials.\cite{kawakami_prl09,yonemitsu_jpsj11b,gomi_jpsj14} The intradimer charge degrees of freedom are studied in reference to anomalous dielectric permittivity,\cite{abdel_prb10} which is associated with polar charge distributions inside dimers.\cite{naka_jpsj10,gomi_prb10,hotta_prb10,dayal_prb11,itoh_prl13} However, the high-frequency charge oscillation mode had not been discussed before Ref.~\citen{iwai_unpub}. Thus, the mechanism and condition for the appearance of this mode are yet to be clarified. 

Here, we show that such a charge oscillation mode emerges in different dimerized systems after the application of a strong pulse of an oscillating electric field. Numerical results are presented in a one-dimensional spinless-fermion ``$t_1$-$t_2$-$V$'' model at half filling and a two-dimensional extended Hubbard model for $\kappa$-(BEDT-TTF)$_2$X at three-quarter filling, which is photoexcited along the $a$- or $c$-axis. The high-frequency charge-oscillation mode is shown to appear in a wide parameter space of ground states with a uniform charge distribution. A close association with time-averaged properties is also revealed. 

\section{Dimerized Models in One and Two Dimensions}
\begin{figure}
\includegraphics[height=13.6cm]{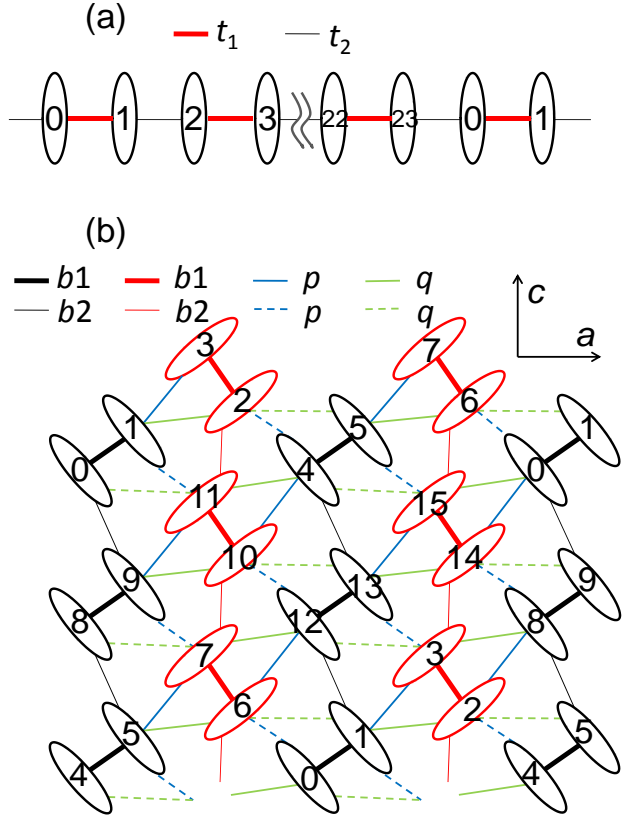}
\caption{(Color online) 
(a) One-dimensional lattice for spinless-fermion model and (b) two-dimensional lattice for $\kappa$-(BEDT-TTF)$_2$X. 
\label{fig:1dedsf_2dkappa_latt}}
\end{figure}

In one dimension, we use one of the simplest models, i.e., a spinless fermion model at half filling, 
\begin{eqnarray}
H_{\mbox{1D}} & = &
t_1 \sum_{n} ( c^\dagger_{2n} c_{2n+1} + c^\dagger_{2n+1} c_{2n}  ) 
+ t_2 \sum_{n} ( c^\dagger_{2n-1} c_{2n} + c^\dagger_{2n} c_{2n-1}  ) \nonumber \\&&
+ V \sum_{j} \left(n_{j}-\frac12 \right) \left(n_{j+1}-\frac12 \right)
\;, \label{eq:1D_model}
\end{eqnarray}
where $ c^\dagger_{j} $ creates a spinless fermion at site $ j $ and $ n_{j} $=$ c^\dagger_{j} c_{j} $. The parameter $ V $ represents the nearest-neighbor repulsion. Large and small transfer integrals, $ t_1 $ and $ t_2 $, are alternated, as shown in Fig.~\ref{fig:1dedsf_2dkappa_latt}(a). 
A 24-site system with a periodic boundary condition is used. The distance between neighboring sites is set to be equal and unity. We use $ t_{2} = -0.1 $ and vary $ t_{1} $ and $ V $. 

In two dimensions, we use an extended Hubbard model at three-quarter filling, 
\begin{eqnarray}
H_{\mbox{2D}} & = & 
\sum_{\langle ij \rangle \sigma}
t_{ij}  ( c^\dagger_{i\sigma} c_{j\sigma} + c^\dagger_{j\sigma} c_{i\sigma}  )
+U\sum_i \left(n_{i\uparrow}-\frac34 \right) 
\left(n_{i\downarrow}-\frac34 \right) \nonumber \\&&
+\sum_{\langle ij \rangle} V_{ij} \left(n_i-\frac32 \right) 
\left(n_j-\frac32 \right)
\;, \label{eq:2D_model}
\end{eqnarray}
where $ c^\dagger_{i\sigma} $ creates an electron with spin $ \sigma $ at site $ i $, $ n_{i\sigma} $=$ c^\dagger_{i\sigma} c_{i\sigma} $, and $ n_i $=$ \sum_\sigma n_{i\sigma} $. The parameter $ U $ represents the on-site Coulomb repulsion. The transfer integral $ t_{ij} $ and the intersite Coulomb repulsion $ V_{ij} $ depend on the bond $ ij $, as shown in Fig.~\ref{fig:1dedsf_2dkappa_latt}(b). A 16-site system with periodic boundary conditions is used. The intermolecular distances and angles are taken from the structural data for $\kappa$-(BEDT-TTF)$_2$Cu[N(CN)$_2$]Cl.\cite{mori_bcsj99,watanabe_sm99} Unless stated otherwise, we use, in units of eV, $ t_{b1} = -0.3006 $,  $ t_{b2} = -0.1148 $,  $ t_{p} = -0.1107 $, and  $ t_{q} =  0.0424 $, which are estimated from the extended H\"uckel calculation,\cite{watanabe_sm99} $ U = 0.8 $, $ V_{b1} = 0.40 $, $ V_{b2} = 0.24 $, $ V_{p} = 0.28 $, and $ V_{q} = 0.24 $. In Eq.~(\ref{eq:2D_model}), the constant term is subtracted in such a way that the total energy becomes zero in equilibrium at infinite temperature. 

The initial state is the ground state obtained by the exact diagonalization method. Photoexcitation is introduced through the Peierls phase 
\begin{equation}
c_{i\sigma}^\dagger c_{j\sigma} \rightarrow
\exp \left[
\frac{ie}{\hbar c} \mbox{\boldmath $r$}_{ij} \cdot \mbox{\boldmath $A$}(t)
\right] c_{i\sigma}^\dagger c_{j\sigma}
\;, \label{eq:photo_excitation}
\end{equation}
which is substituted into Eq.~(\ref{eq:2D_model}) [and its spinless analog is substituted into Eq.~(\ref{eq:1D_model})] for each combination of sites $ i $ and $ j $ with relative position $ \mbox{\boldmath $r$}_{ij}=\mbox{\boldmath $r$}_j-\mbox{\boldmath $r$}_i $. We employ symmetric one-cycle electric-field pulses\cite{yonemitsu_jpsj17a,yonemitsu_jpsj15,yanagiya_jpsj15} and use the time-dependent vector potential 
\begin{equation}
\mbox{\boldmath $A$} (t) = \frac{c\mbox{\boldmath $F$}}{\omega_c} \left[ \cos (\omega_c t)-1 \right] 
\theta (t) \theta \left( \frac{2\pi}{\omega_c}-t \right)
\;, \label{eq:monocycle_pulse}
\end{equation}
where $ \mbox{\boldmath $F$} $ describes the amplitude ($ F = \mid  \mbox{\boldmath $F$} \mid $) and polarization of the electric field. Unless stated otherwise, the central frequency $ \omega_c $ is chosen to be $ \omega_c=0.7 $, which is above the main charge-transfer excitations, as shown below. Hereafter, frequencies $ \omega $ including $ \omega_c $ are also shown in units of eV (with $\hbar$=1). The optical conductivity spectra are calculated for the ground states as before.\cite{yonemitsu_jpsj11b} The time-dependent Schr\"odinger equation is numerically solved by expanding the exponential evolution operator with a time slice $ dt $=0.02 to the 15th order and by checking the conservation of the norm.\cite{yonemitsu_prb09} 

\section{Intradimer and Interdimer Bond Densities in Two-Dimensional Case}
\begin{figure}
\includegraphics[height=13.6cm]{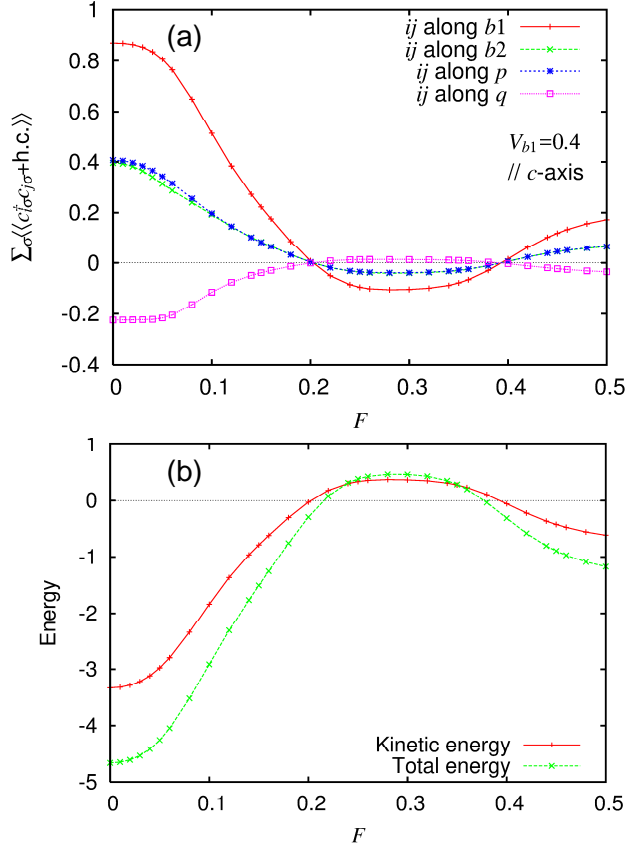}
\caption{(Color online) 
(a) Time-averaged bond densities $ \sum_\sigma \langle \langle \left( c^\dagger_{i\sigma} c_{j\sigma} + c^\dagger_{j\sigma} c_{i\sigma} \right) \rangle \rangle $ for different combinations of $ij$, (b) time-averaged kinetic energy, and total energy after photoexcitation along $c$-axis as functions of field amplitude $ F $. 
\label{fig:bnd_kete_u0p8vb0p4_w0p70fxq90}}
\end{figure}

In this section, we use the two-dimensional model for $\kappa$-(BEDT-TTF)$_2$X. Some quantities time-averaged after photoexcitation along the $c$-axis are shown in Fig.~\ref{fig:bnd_kete_u0p8vb0p4_w0p70fxq90}. 
In this paper, averages are taken over the interval of $ 5T < t < 10T $ with $ T=2\pi/\omega_c $ as before.\cite{yonemitsu_jpsj17a} 
Figure~\ref{fig:bnd_kete_u0p8vb0p4_w0p70fxq90}(a) shows the time-averaged bond densities $ \sum_\sigma \langle \langle \left( c^\dagger_{i\sigma} c_{j\sigma} + c^\dagger_{j\sigma} c_{i\sigma} \right) \rangle \rangle $ for $ij$ along the $ b_1 $, $ b_2 $, $ p $, and $ q $ bonds as functions of the field amplitude $ F $. Here, $ F $ is shown in units of V/\AA. If we take the length scale $ a $=2.6 \AA from the component parallel to the $c$-axis of the intradimer intermolecular relative position in $\kappa$-(BEDT-TTF)$_2$X, $ F $=0.20 corresponds to $ eaF/\hbar \omega_c $=0.74. As $ F $ increases, the time-averaged bond densities decrease in magnitude, simultaneously vanish at $ F $=0.20, and invert their signs. 

This behavior is inconsistent with effective transfer integrals, which are renormalized by the zeroth-order Bessel functions with bond-dependent arguments\cite{nishioka_jpsj14} and vanish at different values of $ F $. This fact suggests that electrons on different bonds are transferred in a concerted manner for large $ F $, which will be discussed in later sections. This behavior is universally observed for large $ F $. For $ \omega_c $=0.9, the time-averaged bond densities simultaneously invert their signs again at $ eaF/\hbar \omega_c $=0.74. For $ \omega_c $=0.5, which is inside the main charge-transfer excitations in the conductivity spectrum, they are substantially suppressed at $ eaF/\hbar \omega_c $=0.74, but they do not invert their signs for larger values of $ F $ (not shown). 

Figure~\ref{fig:bnd_kete_u0p8vb0p4_w0p70fxq90}(b) shows the time-averaged kinetic energy, i.e., the expectation value of the first term in Eq.~(\ref{eq:2D_model}), and the total energy after photoexcitation. A negative-temperature state is realized when the total energy is positive (i.e., larger than that in equilibrium at infinite temperature). Note that a negative-temperature state is formed after the application of a one-cycle pulse of the electric field in different models.\cite{yonemitsu_jpsj15,yanagiya_jpsj15} For $ \omega_c $=0.9, the time-averaged kinetic energy and the total energy behave similarly to those for $ \omega_c $=0.7 here. For $ \omega_c $=0.5, however, they do not invert their signs; thus, a negative-temperature state is not realized. 

\begin{figure}
\includegraphics[height=6.8cm]{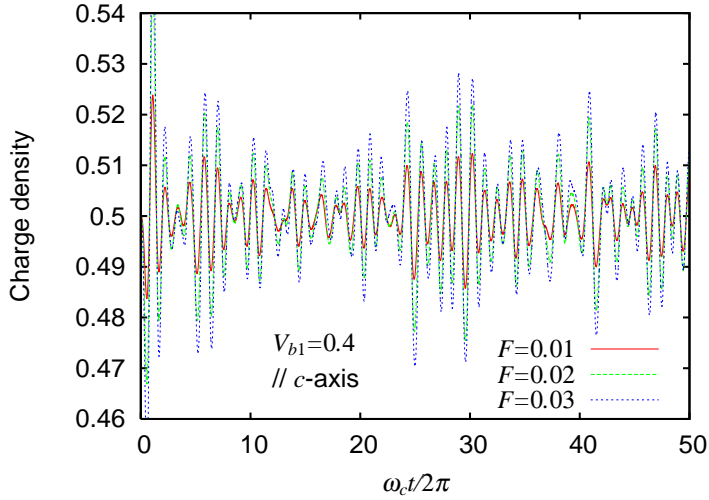}
\caption{(Color online) 
Time evolution of charge density $ 2-\langle n_0 \rangle $ at zeroth site in Fig.~\ref{fig:1dedsf_2dkappa_latt}(b) during and after photoexcitation along $c$-axis with small $ F $. 
\label{fig:dyn_2mne_u0p8vb0p4_w0p70f0p0123q90}}
\end{figure}

To see why the $ F $ dependences of the time-averaged bond densities are not described by the corresponding effective transfer integrals, we need to observe the short-time behavior of transient states. Thus, the time evolution of the charge density $ 2-\langle n_0 \rangle $ is shown in Fig.~\ref{fig:dyn_2mne_u0p8vb0p4_w0p70f0p0123q90} for photoexcitation along the $c$-axis with small values of $ F $. 
For photoexcitation along the $a$-axis or along the $c$-axis, all sites are classified into two groups according to their charge densities for a reason of symmetry. Because the total charge is conserved, the charge distribution among sites is determined once the charge density at one site is known. Consequently, under the condition of Fig.~\ref{fig:dyn_2mne_u0p8vb0p4_w0p70f0p0123q90}, the charge densities at the even-numbered sites in Fig.~\ref{fig:1dedsf_2dkappa_latt}(b) vary with time as shown here, and those at the odd-numbered sites vary in the opposite phase. 

\begin{figure}
\includegraphics[height=6.8cm]{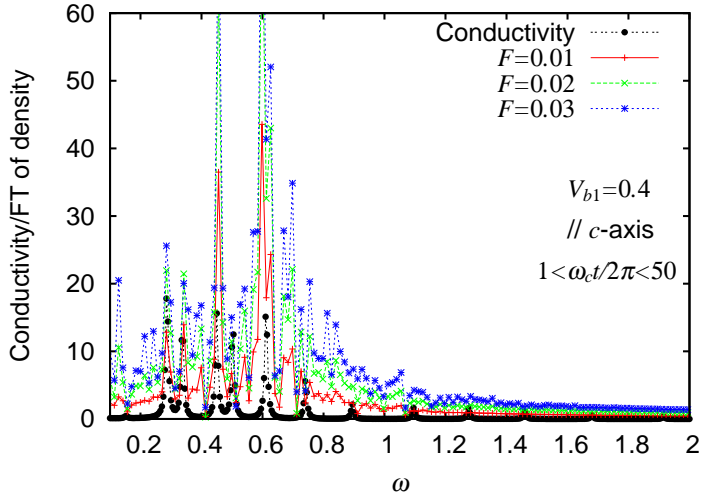}
\caption{(Color online) 
Absolute values of Fourier transforms of time profiles ($ T < t < 50T $) of charge density in Fig.~\ref{fig:dyn_2mne_u0p8vb0p4_w0p70f0p0123q90} compared with optical conductivity spectrum with polarization parallel to $c$-axis in ground state. 
\label{fig:sg_fra_u0p8vb0p4_eps005c_w0p70n1to450fxxxq90}}
\end{figure}

A charge distribution deviating from the initial uniform distribution is caused by photoexcitation. Thus, the time profile of the charge density above is expected to have similar information to the steady-state optical conductivity spectrum. We calculate the absolute values of the Fourier transforms of the time profiles ($ T < t < 50T $ after photoexcitation) of the charge density in Fig.~\ref{fig:dyn_2mne_u0p8vb0p4_w0p70f0p0123q90} and show them in Fig.~\ref{fig:sg_fra_u0p8vb0p4_eps005c_w0p70n1to450fxxxq90}. 
For comparison, we also show the optical conductivity spectrum with polarization parallel to the $c$-axis in the ground state. To facilitate the comparison, we take a rather long time span for the Fourier transforms with a small frequency slice. It is clearly shown that, for small $ F $, the absolute values of the Fourier transforms have peaks with large weights at energies where the conductivity spectrum has peaks. 

The above result shows how charge densities are modulated by photoexcitation and implies that their characteristic time profiles can be observed by optical measurements. In the following sections, we investigate Fourier spectra for large $ F $ more systematically in the one- and two-dimensional models (for specifically polarized fields in the two-dimensional case) where there are two inequivalent sites with respect to charge densities. 

Before discussing specific models, we mention a general fact in noninteracting systems. 
In any noninteracting system, $ H_{\mbox{NI}} = \sum_\lambda \epsilon_\lambda c^\dagger_\lambda c_{\lambda} $, we have $ \left( i d/dt \right) c^\dagger_{\lambda_1} c_{\lambda_2} = \left( \epsilon_{\lambda_2} - \epsilon_{\lambda_1} \right) c^\dagger_{\lambda_1} c_{\lambda_2} $ for any $ \lambda_1 $ and $ \lambda_2 $; thus, the quantities $ Q_{\lambda_1 \lambda_2} \equiv c^\dagger_{\lambda_1} c_{\lambda_2} + c^\dagger_{\lambda_2} c_{\lambda_1} $ and $ P_{\lambda_1 \lambda_2} \equiv -i c^\dagger_{\lambda_1} c_{\lambda_2} +i c^\dagger_{\lambda_2} c_{\lambda_1} $ for $ \lambda_1 \neq \lambda_2 $ oscillate with $ \omega=\mid \epsilon_{\lambda_2} - \epsilon_{\lambda_1} \mid $ similarly to the position and momentum operators of a harmonic oscillator. 

\section{Photoinduced Charge Oscillations in One Dimension \label{sec:1D}}
\begin{figure}
\includegraphics[height=23.6cm]{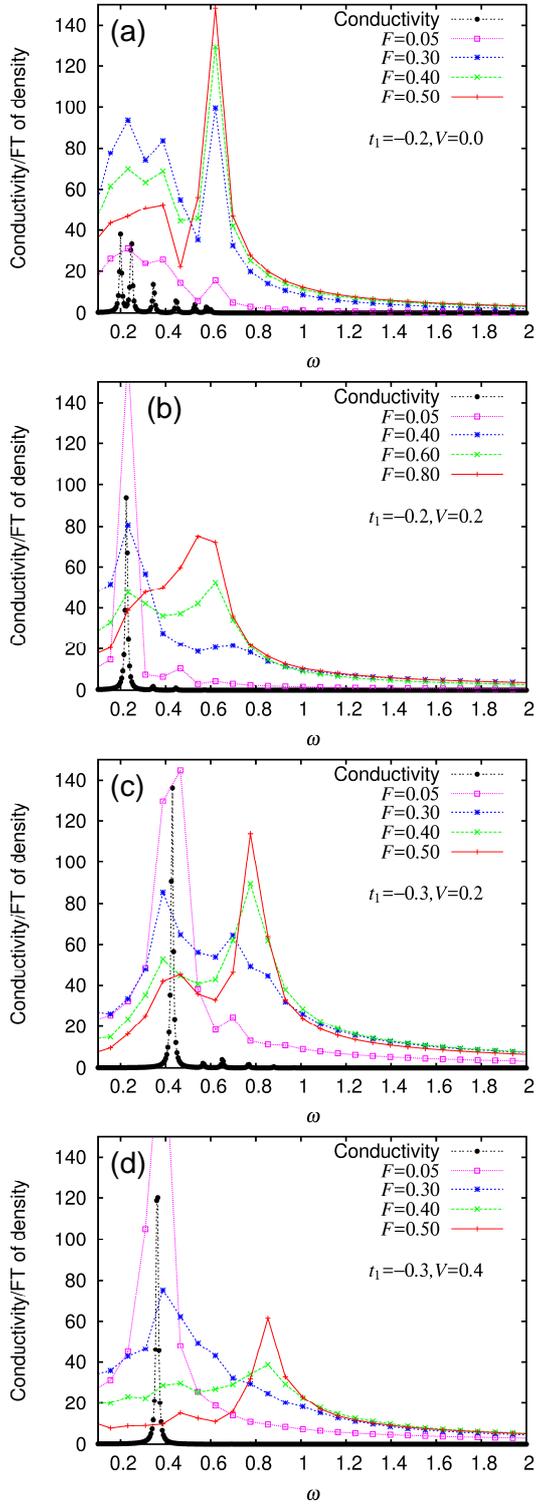}
\caption{(Color online) 
Absolute values of Fourier transforms of time profiles ($ T < t < 10T $) of charge density at any site in Fig.~\ref{fig:1dedsf_2dkappa_latt}(a) with different $F$ and optical conductivity spectrum in ground state for (a) $ t_1 $=$-$0.2, $ V $=0.0, (b) $ t_1 $=$-$0.2, $ V $=0.2, (c) $ t_1 $=$-$0.3, $ V $=0.2, and (d) $ t_1 $=$-$0.3, $ V $=0.4. 
\label{fig:sg_1dedsf_fra_vztoztt1_eps005a_w0p70fxxxq00}}
\end{figure}

To study photoinduced charge oscillations in dimerized systems from a broad perspective, we here use the one-dimensional spinless fermion model in Eq.~(\ref{eq:1D_model}) at half filling. Here and in the following section, we take a shorter time span of $ T < t < 10T $ after photoexcitation for Fourier transforms and refer to the absolute value of the Fourier transform of the time profile of the charge density simply as a Fourier spectrum. For large $ F $ ($ F \geq 0.10$), Fourier spectra are found to be insensitive to the frequency slice. In other words, the broad peak for large $ F $ ($ F \geq 0.10$) is an intrinsic one, except for in the noninteracting case. A larger frequency slice (i.e., a shorter time span) than that used for Fig.~\ref{fig:sg_fra_u0p8vb0p4_eps005c_w0p70n1to450fxxxq90} makes them easy to read. The Fourier spectra thus obtained are shown for different values of $ t_1 $ and $ V $ in Fig.~\ref{fig:sg_1dedsf_fra_vztoztt1_eps005a_w0p70fxxxq00}. 
If the model parameters are in units of eV, $ F $ is shown in units of V/[intersite distance]. 

In the noninteracting case ($ V $=0.0) [Fig.~\ref{fig:sg_1dedsf_fra_vztoztt1_eps005a_w0p70fxxxq00}(a)] and for $ \mid t_1 \mid > \mid t_2 \mid $, the one-electron states of energies from $ -\mid t_1 \mid - \mid t_2 \mid $ to $ -\mid t_1 \mid + \mid t_2 \mid $ are occupied, and those from $ \mid t_1 \mid - \mid t_2 \mid $ to $ \mid t_1 \mid + \mid t_2 \mid $ are unoccupied. Then, the optical conductivity spectrum has weights in the energy range from $ 2(\mid t_1 \mid - \mid t_2 \mid) $ to $ 2(\mid t_1 \mid + \mid t_2 \mid) $ [from 0.2 to 0.6 in Fig.~\ref{fig:sg_1dedsf_fra_vztoztt1_eps005a_w0p70fxxxq00}(a)]. They form discrete peaks owing to the finite-size effect, and their oscillator strengths become larger as the energy is lowered. For small $ F $ ($ F $=0.05), the weights of the Fourier spectrum are in this energy range. As $ F $ increases, the weights first increase as a whole ($ F $=0.30), but the low-energy part is then decreased, while the high-energy part is further increased ($ 0.30 < F < 0.50 $). For large $ F $ ($ F $=0.50), the charge density oscillates with $ \omega_{\mbox{osc}} = 2(\mid t_1 \mid + \mid t_2 \mid) $ (i.e., with a period of $ \pi/(\mid t_1 \mid + \mid t_2 \mid) $). 

In general, on a bipartite lattice (with sublattices ``e'' and ``o'') without interactions, 
$ H_{\mbox{BL}} = \sum_{i,j} t_{i,j} 
\left( c^\dagger_{e,i} c_{o,j} + c^\dagger_{o,j} c_{e,i} \right) $, where all sites are equivalent in terms of the network of transfer integrals, the quantity 
$ Q \equiv \sum_{i,j} \left( c^\dagger_{e,i} c_{e,j} 
- c^\dagger_{o,i} c_{o,j} \right) $ behaves as a harmonic oscillator, 
$ \left( i d/dt \right)^2 Q = \omega^2 Q $, which is derived through the double commutator. In the ground state, its expectation value is zero. 
If such a system has $N$ sites, we define 
$ c_{e,k}=\sqrt{2/N} \sum_{j} e^{ikj} c_{e,j} $ and 
$ c_{o,k}=\sqrt{2/N} \sum_{j} e^{ikj} c_{o,j} $ to have 
$ Q = (N/2) \left( c^\dagger_{e,k=0}c_{e,k=0} - c^\dagger_{o,k=0}c_{o,k=0} \right) $, whose motion is governed by the term 
$ H_{\mbox{BL},k=0} = (t_1+t_2)\left( c^\dagger_{e,k=0} c_{o,k=0} + c^\dagger_{o,k=0} c_{e,k=0} \right) $ in $ H_{\mbox{BL}} = \sum_k H_{\mbox{BL},k} $ in the present case. 
If $ t_1 t_2 < 0 $, we multiply the creation and annihilation operators for sites 2, 3, 6, 7, etc., by $(-1)$ to have $ t_1 t_2 > 0 $ at first, or equivalently we use $ Q = (N/2) \left( c^\dagger_{e,k=\pi}c_{e,k=\pi} - c^\dagger_{o,k=\pi}c_{o,k=\pi} \right) $ and $ H_{\mbox{BL},k=\pi} = (t_1-t_2)\left( c^\dagger_{e,k=\pi} c_{o,k=\pi} + c^\dagger_{o,k=\pi} c_{e,k=\pi} \right) $ instead of the above. Then, it is straightforward to show that $ \omega = \omega_{\mbox{osc}} $ and the charge oscillation is undamped in the noninteracting case. However, with interactions, $ \left( i d/dt \right)^2 Q \neq \omega^2 Q $ and the charge oscillation is damped. 

The above behavior is basically maintained in an interacting case [Fig.~\ref{fig:sg_1dedsf_fra_vztoztt1_eps005a_w0p70fxxxq00}(b)]. In the optical conductivity spectrum, the oscillator strength is concentrated on the low-energy peak owing to the exciton effect. The weight of the Fourier spectrum for small $ F $ ($ F $=0.05) is also concentrated on a peak at the corresponding energy. However, for large $ F $, the low-energy part is decreased and the high-energy part is increased as $ F $ increases ($ 0.40 < F < 0.80 $). The frequency of the strong-field-induced charge oscillation is given by $ \omega_{\mbox{osc}} = 2(\mid t_1 \mid + \mid t_2 \mid) $. Compared with the noninteracting case [Fig.~\ref{fig:sg_1dedsf_fra_vztoztt1_eps005a_w0p70fxxxq00}(a)], where the charge oscillation is undamped owing to integrability, it is damped; thus, the peak height is smaller. 

Numerical results for a different value of $ t_1 $ are shown in Figs.~\ref{fig:sg_1dedsf_fra_vztoztt1_eps005a_w0p70fxxxq00}(c) and \ref{fig:sg_1dedsf_fra_vztoztt1_eps005a_w0p70fxxxq00}(d). The $ F $ dependence of the Fourier spectrum is similar to above, especially for large $ F $ ($ 0.30 < F < 0.50 $). The frequency of the strong-field-induced charge oscillation is given by $ \omega_{\mbox{osc}} = 2(\mid t_1 \mid + \mid t_2 \mid) $ again, which is 0.8 for Figs.~\ref{fig:sg_1dedsf_fra_vztoztt1_eps005a_w0p70fxxxq00}(c) and \ref{fig:sg_1dedsf_fra_vztoztt1_eps005a_w0p70fxxxq00}(d). For $ t_1 $=$-$0.4,  we find a peak at $ \omega_{\mbox{osc}} $=1.0 for $ V $=0.4 and $ V $=0.6 (not shown), which also satisfies the above equation for $ \omega_{\mbox{osc}} $. When we further increase $ \mid t_1 \mid $, we find that it is occasionally necessary to increase the central frequency $ \omega_c $ as well (e.g., $ \omega_c $=1.4 for $t_1 $=$-$0.6 and $t_1 $=$-$0.8 for $ \omega_c $ to resonate with $ \omega_{\mbox{osc}} $) to obtain the strong-field-induced charge oscillation and that it always appears at a frequency near $ \omega_{\mbox{osc}} = 2(\mid t_1 \mid + \mid t_2 \mid) $. This relation is maintained even when the nearest-neighbor repulsion strengths are alternated. This relation is also maintained even away from half filling. This charge oscillation is not observed in the charge-ordered phase with $ V > 2 \mid t_1 \mid $. 

\begin{figure}
\includegraphics[height=13.6cm]{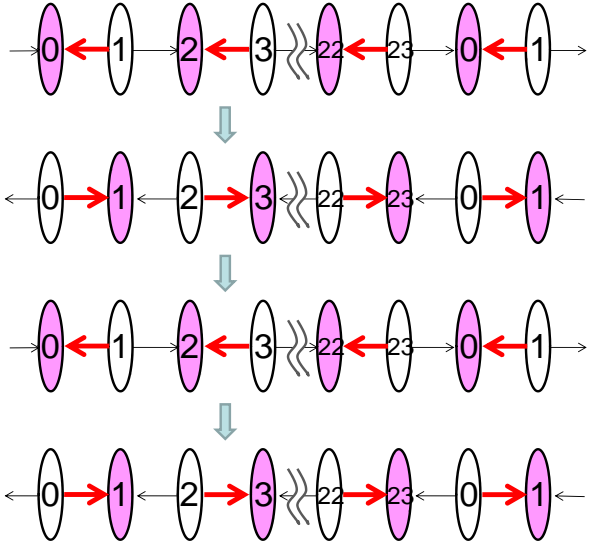}
\caption{(Color online) 
Schematic view of photoinduced charge oscillation. 
\label{fig:1dedsf_osc}}
\end{figure}

The above relation between $ \omega_{\mbox{osc}} $ and the transfer integrals implies that this charge oscillation is realized by simultaneous charge transfers through the $ t_1 $ and $ t_2 $ bonds, as shown in Fig.~\ref{fig:1dedsf_osc}. 
At a site, fermions are simultaneously transferred from the neighboring sites on both sides to this site and back to these two sites. Because these charge-transfer processes are realized coherently everywhere in the system, the whole process is effectively described by a two-site model, $ H_{0} 
= t_{0} \left( c^\dagger_{e} c_{o} + c^\dagger_{o} c_{e} \right) 
= t_{0} \left(c^\dagger_{b} c_{b} - c^\dagger_{a} c_{a} \right) $, where 
$ c_{b(a)} = (c_{e} \pm c_{o} )/\sqrt{2} $ with one fermion 
$ c^\dagger_{e} c_{e} + c^\dagger_{o} c_{o}
= c^\dagger_{b} c_{b} + c^\dagger_{a} c_{a} = 1 $, 
or equivalently by a one-spin model, $ H_{0} = t_{0} \sigma^x $, 
where $ \sigma^x $ is the Pauli matrix with eigenvalues $\pm1$. Its time evolution operator is described (with $\hbar$=1) by 
$ e^{-it H_{0}} = e^{ -itt_{0} \sigma^x } $. 
At $ t = \pi/(2\mid t_{0} \mid) $, the operator 
$ e^{\mp i(\pi/2) \sigma^x} = \mp i \sigma^x $ 
gives a complete charge transfer (spin flip). 
At $ t = \pi/(\mid t_{0} \mid) $, the operator 
$ e^{\mp i\pi \sigma^x} = -1 $ completes a cycle of the charge oscillation. 
These operators are reminiscent of local unitary operators discussed for a discrete time crystal\cite{else_prl16,yao_prl17} in many-body-localized driven systems\cite{zhang_n17,choi_n17}. Here, we do not use a continuous wave but a pulse; thus, we do not need a many-body-localized system to avoid thermalization, which is similar to the situation in Refs.~\citen{yonemitsu_jpsj17b} and \citen{poletti_pra11}. Of course, the pulse-induced charge oscillation decays with time and the Fourier spectra show a broad peak; thus, only the short-time behavior is approximately described by this two-site one-fermion model. 

\section{Photoinduced Charge Oscillations in Two Dimensions}
\begin{figure}
\includegraphics[height=23.6cm]{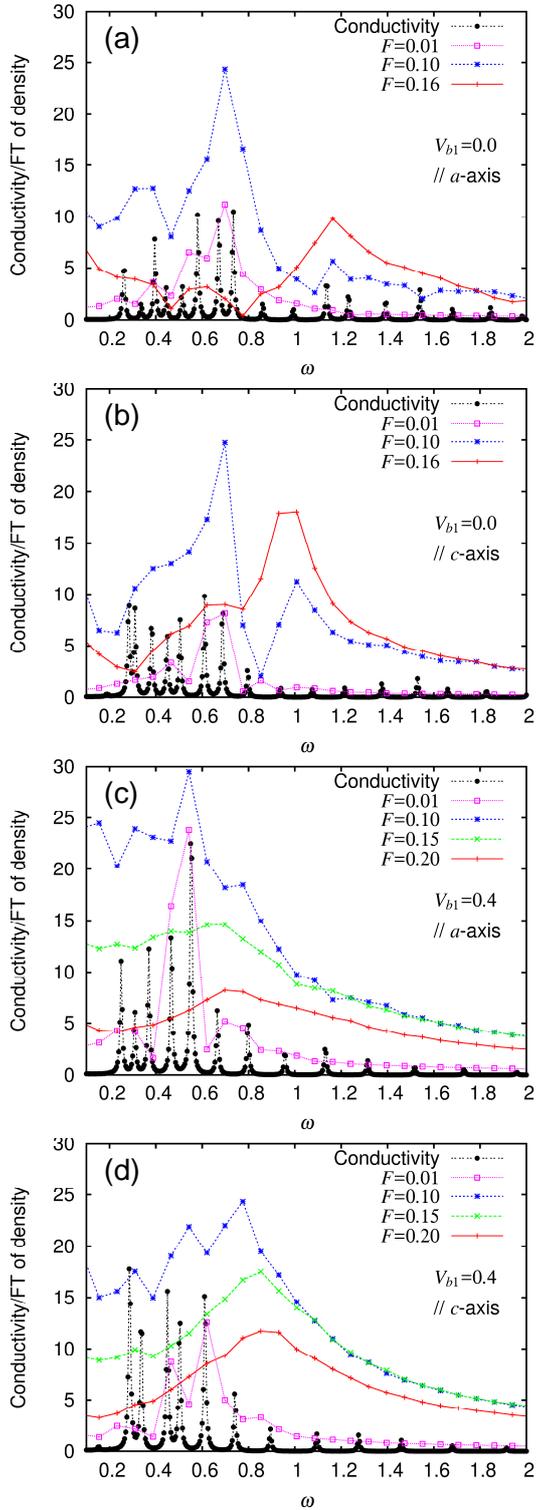}
\caption{(Color online) 
Absolute values of Fourier transforms of time profiles ($ T < t < 10T $) of charge density at any site in Fig.~\ref{fig:1dedsf_2dkappa_latt}(b) with different $F$ and optical conductivity spectrum in ground state for (a), (b) $ V_{b1} $=$ V_{b2} $=$ V_{p} $=$ V_{q} $=0.0 with polarizations along (a) $a$- and (b) $c$-axes, and (c), (d) $ V_{b1} $=0.40, $ V_{b2} $=0.24, $ V_{p} $=0.28, and $ V_{q} $=0.24 with polarizations along (c) $a$- and (d) $c$-axes. In all cases, $ U $=0.8. 
\label{fig:sg_fra_u0p8vbz_eps005y_w0p70fxxxqy}}
\end{figure}

Now, we return to the two-dimensional model for $\kappa$-(BEDT-TTF)$_2$X in Eq.~(\ref{eq:2D_model}) at three-quarter filling and take the time span of $ T < t < 10T $ again for Fourier transforms. The Fourier spectra thus obtained are shown for different strengths of intersite repulsive interactions and polarizations of photoexcitation in Fig.~\ref{fig:sg_fra_u0p8vbz_eps005y_w0p70fxxxqy}. 

In the case without intersite repulsion ($ V_{ij} $=0.0), the Fourier spectra are shown in  Fig.~\ref{fig:sg_fra_u0p8vbz_eps005y_w0p70fxxxqy}(a) [Fig.~\ref{fig:sg_fra_u0p8vbz_eps005y_w0p70fxxxqy}(b)] for polarization along the $a$- ($c$-)axis. For small $ F $ ($ F $=0.01), their weights are mainly distributed in the energy range $ \omega < 0.8 $, similarly to the corresponding conductivity spectra. As $ F $ increases, the weights first increase as a whole ($ F $=0.10), but the low-energy part ($ \omega < 0.8 $) is then decreased, while the high-energy part is further increased ($ 0.10 < F < 0.16 $). The peak energy for large $ F $ ($ F $=0.16) appears around $ \omega $=1.2 for polarization along the $a$-axis [Fig.~\ref{fig:sg_fra_u0p8vbz_eps005y_w0p70fxxxqy}(a)], which is slightly lower than $ \omega_{\mbox{osc}} = 2(\mid t_{b1} \mid + \mid t_{b2} \mid +2 \mid t_p \mid) $, amounting to 1.27 here, and around $ \omega $=1.0 for polarization along the $c$-axis [Fig.~\ref{fig:sg_fra_u0p8vbz_eps005y_w0p70fxxxqy}(b)], which is close to $ \omega_{\mbox{osc}} = 2(\mid t_{b1} \mid + \mid t_{b2} \mid +2 \mid t_q \mid) $, amounting to 1.00 here. The peak energy for large $ F $ generally well matches the equation for $ \omega_{\mbox{osc}} $ above for each polarization in cases with different values of $ t_{ij} $ and $ U $ as long as the intersite interactions are absent, $ V_{ij} $=0.0. 

\begin{figure}
\includegraphics[height=16.0cm]{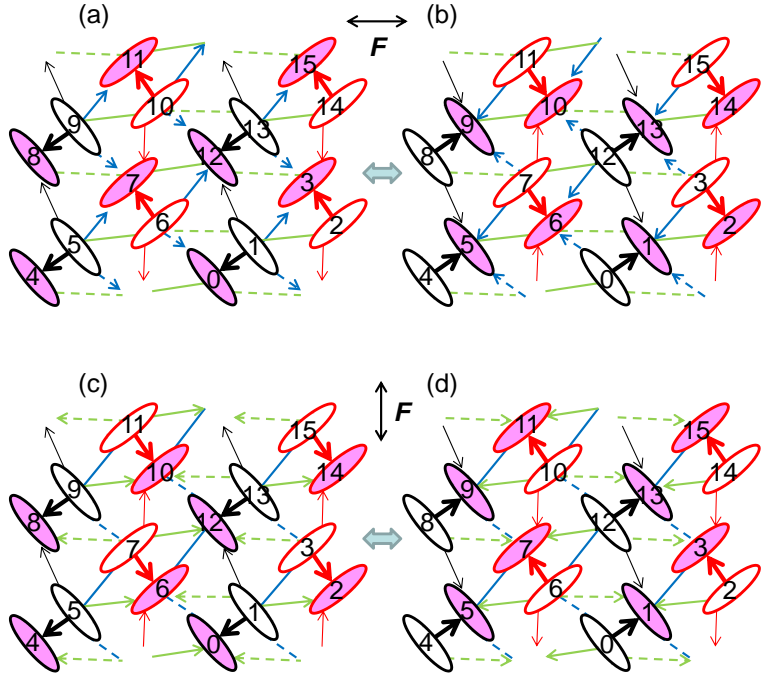}
\caption{(Color online) 
Schematic view of charge oscillation between (a) and (b) after photoexcitation along $a$-axis and that between (c) and (d) after photoexcitation along $c$-axis. 
\label{fig:2dkappa_osc}}
\end{figure}

The above relation between $ \omega_{\mbox{osc}} $ and the transfer integrals is similar to that in the one-dimensional spinless fermion model. The strong-field-induced charge oscillation is realized by simultaneous charge transfers through the $ t_{b1} $, $ t_{b2} $, and two $ t_p $ bonds for polarization along the $a$-axis, as shown in Figs.~\ref{fig:2dkappa_osc}(a) and \ref{fig:2dkappa_osc}(b), and those through the $ t_{b1} $, $ t_{b2} $, and two $ t_q $ bonds for polarization along the $c$-axis, as shown in Figs.~\ref{fig:2dkappa_osc}(c) and \ref{fig:2dkappa_osc}(d). 
Through these bonds, the charge-rich sites are connected to the charge-poor sites. At a charge-poor site, holes are simultaneously transferred from all of the four neighboring charge-rich sites to this site and then back to these four sites. Thus, this oscillation looks like an electronic breathing mode or a checkerboard pattern where black and white are time-periodically exchanged. 

The fact that $ \omega_{\mbox{osc}} $ is independent of the sign of $ t_p $ or that of $ t_q $ is due to the particular structure shown in Fig.~\ref{fig:1dedsf_2dkappa_latt}(b). If we multiply the creation and annihilation operators for red dimers (sites 2, 3, 6, 7, 10, 11, 14, and 15) by $(-1)$, it corresponds to inverting the signs of $ t_p $ and $ t_q $. In general, the equation for $ \omega_{\mbox{osc}} $ is sensitive to the sign of the transfer integral. For instance, if we add a third-neighbor hopping term, $ t_4 \sum_j \left( c^\dagger_{j} c_{j+3} + c^\dagger_{j+3} c_{j} \right) $, to the spinless model of Eq.~(\ref{eq:1D_model}), $ \omega_{\mbox{osc}} $ depends on the relative signs of the transfer integrals and it does not always correspond to the highest-energy peak of the optical conductivity spectrum in the noninteracting case: $ \omega_{\mbox{osc}} = 2 \mid t_1 + t_2 + 2 t_4 \mid $ if $ t_1 t_2 > 0 $ and $ \omega_{\mbox{osc}} = 2 \mid t_1 - t_2 \mid $ if $ t_1 t_2 < 0 $ irrespective of a second-neighbor hopping term, at least for $ \mid t_1 \mid , \mid t_2 \mid \geq 2 \mid t_4 \mid $. 

The above behaviors for large $ F $ are partly changed by intersite repulsive interactions. Examples are shown in Fig.~\ref{fig:sg_fra_u0p8vbz_eps005y_w0p70fxxxqy}(c) [Fig.~\ref{fig:sg_fra_u0p8vbz_eps005y_w0p70fxxxqy}(d)] for polarization along the $a$- ($c$-)axis. The parameters used for Fig.~\ref{fig:sg_fra_u0p8vbz_eps005y_w0p70fxxxqy}(d) are the same as those used for Fig.~\ref{fig:bnd_kete_u0p8vb0p4_w0p70fxq90}. For small $ F $ ($ F $=0.01), their weights are again mainly distributed in the energy range $ \omega < 0.8 $, similarly to the corresponding conductivity spectra. As $ F $ increases, the weights first increase as a whole (with a maximum around $ F $=0.10). These behaviors so far are similar to those in the case without intersite repulsion. However, as $ F $ further increases, the low-energy part is significantly decreased, and the high-energy part is also decreased to some degree ($ 0.10 < F< 0.20 $). As a consequence, a very broad peak is formed in the spectra. For both polarizations, the peak energies are significantly lower than $ \omega_{\mbox{osc}} $ in the case without intersite repulsion. The intersite repulsive interactions (attractive interactions between charge-rich and charge-poor sites) slow down this charge oscillation. Even in one dimension, if we add a second-neighbor hopping term to the spinless model of Eq.~(\ref{eq:1D_model}) to allow an exchange of fermions, the frequency of the strong-field-induced charge oscillation is altered by interactions (not shown). 

Thus, immediately after the application of a strong pulse of an oscillating electric field, charge densities coherently oscillate with a finite lifetime. This fact is consistent with the finding in Fig.~\ref{fig:bnd_kete_u0p8vb0p4_w0p70fxq90}(a), that is, the time-averaged bond densities are synchronized with each other as functions of $ F $. During the charge oscillation, holes are transferred through different bonds simultaneously with a common frequency; thus, the time-averaged bond densities on different bonds behave similarly as functions of $ F $. These simultaneous charge transfers are necessary to suppress the rise in the entropy and to realize a negative-temperature state in Fig.~\ref{fig:bnd_kete_u0p8vb0p4_w0p70fxq90}(b). 

\begin{figure}
\includegraphics[height=20.4cm]{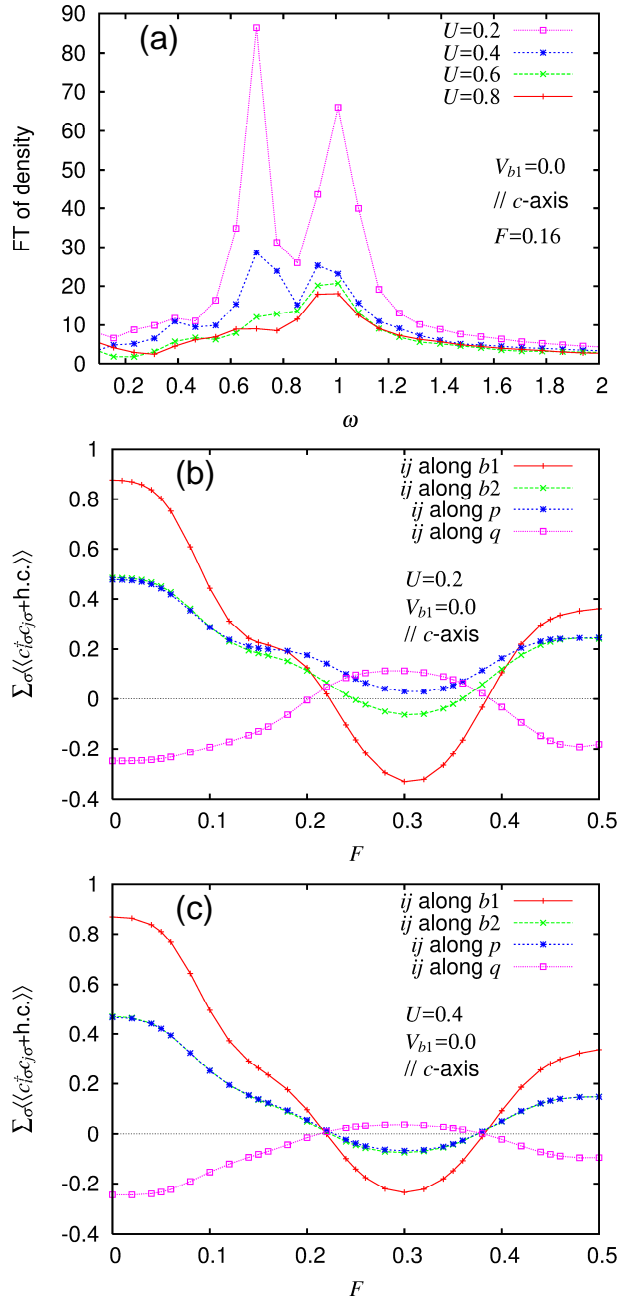}
\caption{(Color online) 
(a) Fourier spectra similar to Fig.~\ref{fig:sg_fra_u0p8vbz_eps005y_w0p70fxxxqy} but for $ F $=0.16 and different $ U $, and (b),(c) time-averaged bond densities similar to Fig.~\ref{fig:bnd_kete_u0p8vb0p4_w0p70fxq90}(a) but for (b) $ U $=0.2 and (c) $ U $=0.4. The parameters used here are $ V_{b1} $=$ V_{b2} $=$ V_{p} $=$ V_{q} $=0.0, and the polarization is along the $c$-axis. 
\label{fig:fra_bnd2_uxvb0p0_w0p70q90}}
\end{figure}

Now, we consider what is necessary for this synchronized behavior. Each of the effective transfer integrals is renormalized by the zeroth-order Bessel function with the bond-dependent argument proportional to the inner product of $ \mbox{\boldmath $F$} $ and $ \mbox{\boldmath $r$}_{ij} $. If the effective transfer integrals governed the long-time behavior, the synchronized behavior would not be obtained. The picture based on the effective transfer integrals should basically be applied to noninteracting and weakly interacting systems. Then, in the case without intersite repulsion and with different values of $ U $, we show Fourier spectra for large $ F $ ($ F $=0.16) and polarization along the $c$-axis [as used in Fig.~\ref{fig:sg_fra_u0p8vbz_eps005y_w0p70fxxxqy}(b)] in Fig.~\ref{fig:fra_bnd2_uxvb0p0_w0p70q90}(a). 
In general, charge oscillations are undamped in the noninteracting case; thus, their lifetimes are long for weakly interacting systems. That is why, as $ U $ decreases, the overall weights in the Fourier spectra increase. Roughly speaking, these spectra consist of a low-energy part ($ \omega < 0.8 $) corresponding to the conductivity spectrum and a high-energy peak around $ \omega $=1.0 due to the strong-field-induced charge oscillation. As $ U $ decreases, the suppression of the low-energy part becomes weak; thus, the strong-field-induced charge oscillation becomes relatively weaker. This implies that the tendency for simultaneous charge transfers becomes weak as $ U $ decreases. 

Then, we show the time-averaged bond densities $ \sum_\sigma \langle \langle \left( c^\dagger_{i\sigma} c_{j\sigma} + c^\dagger_{j\sigma} c_{i\sigma} \right) \rangle \rangle $ in Figs.~\ref{fig:fra_bnd2_uxvb0p0_w0p70q90}(b) and \ref{fig:fra_bnd2_uxvb0p0_w0p70q90}(c) for the small-$ U $ cases of Fig.~\ref{fig:fra_bnd2_uxvb0p0_w0p70q90}(a). Indeed, for small $ U $ ($ U $=0.2), the time-averaged bond densities are not synchronized as functions of $ F $ [Fig.~\ref{fig:fra_bnd2_uxvb0p0_w0p70q90}(b)]. As $ F $ increases, the effective transfer integral for the $ b_2 $ bond first vanishes, then that for the $ p $ bond vanishes, and so on. Thus, the behaviors of the time-averaged bond densities are not simply described by the corresponding effective transfer integrals. For larger $ U $, the time-averaged bond densities are almost [Fig.~\ref{fig:fra_bnd2_uxvb0p0_w0p70q90}(c) for $ U $=0.4] or fully [as in Fig.~\ref{fig:bnd_kete_u0p8vb0p4_w0p70fxq90}(a) for $ U $=0.6 (not shown)] synchronized as functions of $ F $. Thus, interactions are found to be essential for the simultaneous charge transfers: the interactions cause damping of charge oscillations and synchronization of the charge transfers at the same time, which result in the strong damping of low-frequency charge oscillations and the relatively weak damping of the high-frequency charge oscillation. This fact is reminiscent of the situation in a discrete time crystal, where interactions are essential for collective synchronization in strongly disordered systems.\cite{yao_prl17} 

The emergence of the strong-field-induced charge oscillation is basically limited to the polarizations along the $a$- and $c$-axes, which guarantee that there are only two groups of sites according to their charge densities and that the Fourier spectra are common to all sites. When the polarization deviates from these axes, the charge densities at the four sites in the unit cell become nonequivalent after photoexcitation; thus, the time profiles depend on the site and their Fourier spectra become different. Even for large $ F $, substantially large weights remain at low energies. 

Thus, the presence of only two nonequivalent sites with respect to charge density is important. Dimerized structures are favorable in this sense. Indeed, similar results are obtained in a one-dimensional spin-1/2 model at three-quarter filling with a similar degree of dimerization, where the frequency of the strong-field-induced charge oscillation is given by $ \omega_{\mbox{osc}} = 2(\mid t_1 \mid + \mid t_2 \mid) $ for different values of $ U $ and $ V $ in the uniform-charge-density phase. 

\section{Conclusions and Discussion}
Instead of the long-time behavior often explained by concepts such as dynamical localization and modified effective interactions, we pay attention to a short-time behavior that is nonlinear with respect to the field amplitude $ F $ and varies on a time scale of the period of the external field. This study is motivated by a recent experiment suggesting the importance of short-time behavior.\cite{iwai_unpub} Because of the dimerized structures and specifically polarized fields considered in this study, all sites are classified into two groups according to their charge densities. Charge densities take a common value within a group after photoexcitation. Numerical results are presented for a one-dimensional spinless-fermion model at half filling and a two-dimensional model for $\kappa$-(BEDT-TTF)$_2$X at three-quarter filling, but the main conclusions are not limited to these models as long as all sites are classified into two groups and the ground state is in the uniform-charge-density phase. 

For small $ F $, Fourier spectra for charge-density time profiles after photoexcitation have peaks at energies where the corresponding conductivity spectra have peaks. For large $ F $, the spectral distribution is changed and has a peak at a single energy on the high-energy side. For the models we use in this paper and without intersite repulsion in two dimensions, the peak energy is given by twice the sum of the absolute values of the transfer integrals between a site and all neighboring sites with different charge density. In two dimensions, this peak energy is lowered by intersite repulsion. However, we can construct a model where the long-range hopping increases, decreases, or maintains the peak energy depending on the relative signs of transfer integrals. This field-induced charge oscillation appears only when $ F $ is large; thus, it is a nonlinear phenomenon that emerges when a strong pulse of an oscillating electric field is applied to dimerized systems and charge densities are shaken coherently. This strong-field-induced charge oscillation is considered to be closely related to the newly observed reflectivity peak in photoexcited $\kappa$-(BEDT-TTF)$_2$Cu[N(CN)$_2$]Br on the high-energy side of the main reflectivity spectrum.\cite{iwai_unpub}

The coherence associated with the strong-field-induced charge oscillation is responsible for the behaviors of the time-averaged bond densities, which decrease in magnitude, simultaneously vanish at a particular value of $ F $, and invert their signs as $ F $ increases. After the sign inversion, a negative-temperature state is realized, which implies that the rise in the entropy is suppressed. This suppression is enabled by coherently shaking charge densities. Note that a negative-temperature state is more generally realized even without dimerization.\cite{yanagiya_jpsj15} 

For continuous waves, thermalization is suppressed in a many-body-localized system, which is necessary to realize a discrete time crystal.\cite{else_prl16,yao_prl17} The time evolution operators for the two-site one-fermion model that is referred to for the discussion of the frequency of the strong-field-induced charge oscillation in Sect.~\ref{sec:1D} are similar to local unitary operators discussed for a discrete time crystal. This similarity may be helpful when considering the possibility of emergent charge oscillations in different situations. 

\begin{acknowledgment}
The author is grateful to S. Iwai and Y. Tanaka for various discussions. 
This work was supported by Grants-in-Aid for Scientific Research (C) (Grant No. 16K05459) and Scientific Research (A) (Grant No. 15H02100) from the Ministry of Education, Culture, Sports, Science and Technology of Japan. 
\end{acknowledgment}

\bibliography{68580}

\end{document}